\documentclass[twocolumn,showpacs,preprintnumbers,amsmath,amssymb]{revtex4}

\usepackage{graphicx}
\usepackage{dcolumn}
\usepackage{bm}
\newcommand{\bra}[1]{\langle #1|}
\newcommand{\ket}[1]{|#1\rangle}

\begin{document}

\title{Proposed strategy to sort
semiconducting nanotubes by radius and chirality}

\author{V Narayan}

\affiliation{Department of Physics, Link\"{o}ping University,
SE-581 83, Link\"{o}ping, Sweden}

\date{\today}
\begin{abstract}
We propose a strategy that uses a tunable laser and an alternating
non-linear potential to sort a suspension of assorted semiconducting
nanotubes. Since, a polarized exciton is a dipole, the excited
nanotubes will experience a net force and may then diffuse towards an
electrode. The calculated exciton binding energy suddenly drops to
zero and the force on the nanotube increases dramatically when the
exciton disassociates as the nanotube moves towards the electrode.
The quantum adiabatic theorem  shows that excitons will be polarized
for potential frequencies typical for experiments $\approx 10$~MHz.

\end{abstract}
\maketitle

The unexpected discovery of another state of carbon in single walled
carbon nanotubes\cite{saitobk} (SWNTs) has  attracted a great deal of
theoretical and technological interest. SWNTs can be semiconducting or
metallic\cite{saitobk} depending on their chirality and diameter, and
may be combined to create exotic devices.  Typically an assortment of
SWNTs bound by strong van-der-waals forces are created, thus
individual SWNTs are hard to manipulate.  Recently, SWNT ropes were
seperated in water (containing a surfactant) by sound waves and
centrifugation\cite{smalley}. The resulting suspension contained SWNTs
with an average diameter and length of $0.7$~nm and $130$~nm
respectively. Metallic and semiconducting SWNTs have subsequently been
separated\cite{Krupke} by alternating current (AC)
dielectrophoresis\cite{Pohlbk}. This paper proposes theoretically a
strategy for sorting the remaining semiconducting SWNTs according to
radius and chirality.
 
An electrode defines an alternating exponentially decaying potential
in the suspension, and laser pulses excite electron-hole ($e-h$)
pairs. Since the band gap conveniently scales with the inverse of the
diameter (where a diameter between  $0.6$ to $1.6$~nm corresponds to
$0.4-1.0$~eV) the laser  frequency can be tuned to select some
nanotubes\cite{ichida1}. The external potential frequency is ``slow''
so that the $e-h$ hole pairs are polarized adiabatically, but also
``fast'' enough to prevent the build up of a surface charge and its
associated electrophoretic force. The $e-h$ pairs are expected to
initially screen the external potential \cite{gulbinas2} which will
recover as the exciton density decays. The remaining excitons, as in
the p-n junction of a nanotube solar cell\cite{Freitag}, shall
disassociate and migrate to opposite ends.  The excitonic dipole
experiences a net time average force and the host nanotube may then
diffuse towards the electrode, whereas the inert nanotubes will remain
in place; this is the essence of the separation strategy which can be
classified as light-induced AC dielectrophoresis. The exciton is
considered quantum mechanically to calculate the force on a bare
nanotube, and its dipole moment to estimate the interaction with the
medium.

The time dependent Hamiltonian for an electron and hole on the surface
of a nanotube with positions $r_1$ and $r_2$ with masses $m_1$ and
$m_2$ has the form
\begin{equation}
H=p_{r_1}^{2}/2m_{1}+p_{r_1}^{2}/2m_{1}+V_{C}\left(r_{1},r_{2}\right)+
V_{ext}
\end{equation}
where the external potential
$V_{ext}=e(t/T)(V_{0}e^{-kz_1}-V_{0}e^{-kz_2})$, $V_c$ is the
attractive Coulomb potential, and $V_0$ and $k$ are constants
determined by the gate bias and geometry. Within one cycle the external
potential increases adiabatically from zero to its full value for
$0<t<T$ during the polarization time $T$. The adiabatic regime implies,
that instantaneous eigenstates can be calculated using an effective
time independent Hamiltonian.

In an effort to illustrate the physical trends of the system we have
calculated the adiabatic exciton ground state using a simple Hartree
model by assuming a separable wavefuction
$\Phi(r_{1},r_{2})=\Phi(r_{1})\Phi(r_{2})$  and neglecting
bandstructure effects\cite{kostov,Pedersen2003}. Since the nanotube
length is much greater than its diameter\cite{Pedersen,Ogawa} the
wavefunction in the radial direction can be assumed to be dominated by
directions strong confinement. The radial kinetic energy is assumed to
be much larger than the electron-hole interaction potential since the
exciton radius will be much larger than the carbon bond
length. Mathematically this suggests that we can assume that the
electron and hole wavefunctions are separable in $z$ and $\theta$
coordinates, reducing this 2D problem\cite{janssens} to a 1D
problem. The ground state wavefunction for a nanotube with radius $r$
in the radial direction has the form
$\Psi(\theta)=e^{-i\lambda\theta}{(2\pi)}^{-1/2}$ where the energy
$\lambda=\hbar^2/(2mr^2)$. The Hartree equations expressed in
cylindrical polar coordinates are then, 

\begin{eqnarray}
\left[
{p_{i}^{2}\over{2m_i}}
+ V{ij}+V_{ext}(z_{i})
\right]
\Psi\left(\theta_{i}\right)
\Psi_{i}\left(\bf{z_{i}}\right)  
=E_{i} 
\Psi\left(\theta_{i}\right)
\Psi\left(\bf{z_{i}}\right)
\nonumber \\
V{ij}\left(z_{i},\theta_{i}\right)=
\int\limits_{0}^{\infty} dz_{j}
\int\limits_{0}^{2\pi} d \theta_{j}
\Psi(z_{j})^2 
{1\over{8\pi^{2}\epsilon}}
{1\over{\sqrt{d}}}
\nonumber \\
i,j=e,h \nonumber \\
j\neq i
\end{eqnarray}
where $V_{ij}$ is the electron-hole Coulomb potential,
$d=r^2(2-\cos(\theta_{i}-\theta_{j}))+(z_{i}-z_{j})^2$,
and $\epsilon$ is the dielectric constant.
Multiplying by $e^{i\lambda\theta_{i}}{(2\pi)}^{-1/2}$ and
integrating over $\theta$ coordinates we get,

\begin{equation}
\left[
{-\hbar^2\over{2m}}{\partial\over{\partial^2 z}}
+V_{i}(z_{i})+V_{ext}
\right]
\Psi(z_{i})
=\left[
E_{i}
-{\hbar^2\over{2mr^2}}
\right]\Psi(z_{i})
\end{equation}
and $V_{i}(z_{i})=\int V_{eff}(z_{i}-z_{j}) \Psi(z_{j}) dz_{j}$
where an effective potential $V_{eff}$ is defined
as,
\begin{equation}
V_{eff}(z_{i}-z_{j})={1\over{4\pi^2}} \int 
{1\over{4\pi\epsilon}}
{1\over{\sqrt{d}}}
d\theta_{i}
d\theta_{j}
\end{equation}
Since the effective potential has no analytic form we follow previous
works\cite{jan,Pedersen} and approximate the effective potential to
a simple form
$V_{eff}\approx U_{eff}=1/(4 \pi \epsilon)
[\sigma^2-(z_{i}-z_{j})^2]^{-1/2}$, where $\sigma$ is
a parameter adjusted to make $U_{eff}$ resemble the numerically
integrated average potential $V_{eff}$. The Hartree equations are then
discretized and solved self consistently.

We assume throughout that the torque generated by the external
potential has aligned the nanotube. We have calculated the exciton
binding energy, the dipole moment $p=e|\bra{\Psi} z\ket{\Psi}|$ and the
instantaneous acceleration $a$ experienced by a nanotube with length
$L$ whose left edge is a distance $d$ from the origin. The
acceleration $a$ is related to the electron $\rho_{e}$ and hole
$\rho_{h}$ charge densities,
\begin{equation}
a(z,L)
={{1}\over{\mu L}}
\int_{d}^{d+L}
kV_{ext}(z)
\left[
\rho_{e}(z)-\rho_{h}(z)
\right]
dz
\end{equation}
where the nanotubes mass $m_{t}=\mu L$ where $\mu$ is a constant 
depending on the number of atoms per unit cell.

We consider a typical semiconducting SWNT the $(12,0)$ with diameter
$0.994$~nm and a unit cell of length $0.85$~nm\cite{Lehtinen} which
contains $96$ carbon atoms. We have set $m_{e}=m_{h}=0.08 m_{0}$
(unless stated otherwise), a typical value for a small diameter
semiconducting SWNT\cite{Pedersen}, $\epsilon=3.3$\cite{Pedersen}, and
$\sigma=0.45$~nm. The results are presented for an external potential
characterized by $k=0.001$ in atomic units, and $V_{0}=2.0$~V.

Fig1 shows the binding energy (BE) as a function of $d$ and $t$ for
$L=20-50$~nm. Since the electron-hole attraction is strong in
nanotubes which are approximate 1D system\cite{Spataru}, the BE is
relatively large $\approx 0.6$~eV when the nanotube is far from the
electrode. The BE drops suddenly to zero at a position dependent
critical time $t_c(d)$ indicating a structural transition in the
excitons wavefunction. The exciton BE depends on the balance between
the attractive Coulomb interaction and the potential drop along the
nanotube. The exciton is clearly harder to polarize as the nanotube
length decreases since the potential drop along the nanotube is less
and the electron and hole are restricted to within a smaller
volume. The forces involved: the Coulomb attraction and the external
potential are all non-linear, which explains the non-linear relation
between $d_c(t)$ with the length $L$.

Fig~\ref{fig2} shows the structural transition of the  excitons
wavefunction for $t/T=1.0$ as a nanotube with length $L=30$~nm is
moved towards the gate.  The balance between the polarizing effect of
the external potential and the Coulomb attraction between the electron
and hole determines the exciton's structure. The exciton is strongly
bound when the nanotube is far from the gate and resides at the the
end which is furtherest away from the gate. As the nanotube is moved
closer, the electron charge density piles towards the end closer to
the electrode and drags the hole along. Finally, around $d_c$ the
electron and hole are located at opposite ends of the nanotube and the
exciton has disassociated.

We have assumed so far that the exciton has been polarized
adiabatically.  The adiabatic characteristic time $T_{a}$ can be
determined from the adiabatic theorem which requires an estimate of
the exciton energy spectra.  Specifically, we need to calculate the
two lowest exciton states $\ket{\Phi_{\alpha}}$ and
$\ket{\Phi_{\beta}}$ with energies $E_{\alpha}$ and
$E_{\beta}$. Typically, the conduction and valence band of a
semiconducting nanotube contains degenerate bands. We calculate the
ground state by assuming that $m_{e}=m_{h}=0.08 m_{0}$ and for the
first excited state the electron has occupied the light band with
$m_{e}=0.05 m_{0}$\cite{zhao}.  The adiabatic theorem
imposes\cite{Martins},
\begin{equation}
T \gg T_{a}= \hbar\Gamma{max}/g^{2}_{min}
\end{equation}
where $g_{min}=min_{0<t<T}\left[E_{\alpha}(t)-E_{\beta}(t)\right]$ and
$\Gamma_{max}=max_{0<t<T} |\bra{\Phi_\alpha}  eV_{0}(e^{-k z_{1}}-e^{k
z_{2}}) \ket{\phi_\beta} |$. Fig~\ref{fig3} shows the time dependence
of the exciton binding energies of these states for a nanotube with
length $L=50$~nm and $d=25$~nm. Evidently, after a critical time 
the external potential is able to disassociate the exciton and the
binding energy drops to zero. The higher energy state $\beta$ with
more kinetic energy, is marginally more easier to disassociate. For
these parameters $g_{min}=1.37$~meV and $T_{a}=4.9$~ps. Since  the
bands in the conduction and/or valence band(s) are  often separated by
an energy gap of the order of $0.1$~eV\cite{reich}, $T_{a}=4.9$~ps
represents a deliberately high estimate.  The external potential in
experiments usually has a frequency $10$~MHz corresponding to $T
\approx 10^{-8}$~s thus excitons  will be polarized, though thermal
effects will have to included to determine the charge distribution.

Typically the dielectric constant ($\epsilon<5$) for  semiconducting
nanotubes is small compared to that of water
$\epsilon_{w}=80.0$. These values when inserted into the expression
for the dielectrophoretic force on a sphere indicates\cite{Krupke}
negative dielectrophoresis i.e. no attraction to the electrode.  The
metal nanotubes in contrast have a large dielectric
constant\cite{benedict} (perhaps infinite) in comparison to that of
water and subsequently experience positive dielectrophoresis and are
strongly attracted to the electrode\cite{Krupke}. The dielectric
constant of an excited nanotube can be estimated by considering the
contribution to the dipole moment\cite{bianchetti} from the
exciton\cite{warburton} as shown in Fig~\ref{fig4}. The exciton dipole
moment  even for modest electric fields is much bigger than that of a
water molecule ($6.19\times 10^{-30}$~Cm) since an exciton is a
delocalized and weakly bound entity in comparison to a molecule. The
dipole moment increases dramatically when the exciton disassociates,
so the polarizabilty is greatest during this transition. The
dielectric constant of  the nanotube cannot be determined using the
Clausius-Mossoti relation\cite{ortuno} since this is only valid for a
weakly polarizable material. A large dielectric constant is indicated
here however, since a substantial cancelation of the applied field by
the mobile charges is expected, so an excited nanotube is expected
undergo positive dielectrophoresis. We note that highly polarizable
materials have enormous dielectric constants: living organisms such as
yeast\cite{prodan} ($\epsilon>10^{5}$) contain mobile charges which
can screen the applied field, and a perovskite related
oxide\cite{homes} which is suspected to
contain responsive dipoles also has large dielectric constant.
We expect excited nanotubes to have a similarly large dielectric
constant.

\begin{figure}[t!]
\begin{center}
\includegraphics[width=8cm,angle=0]{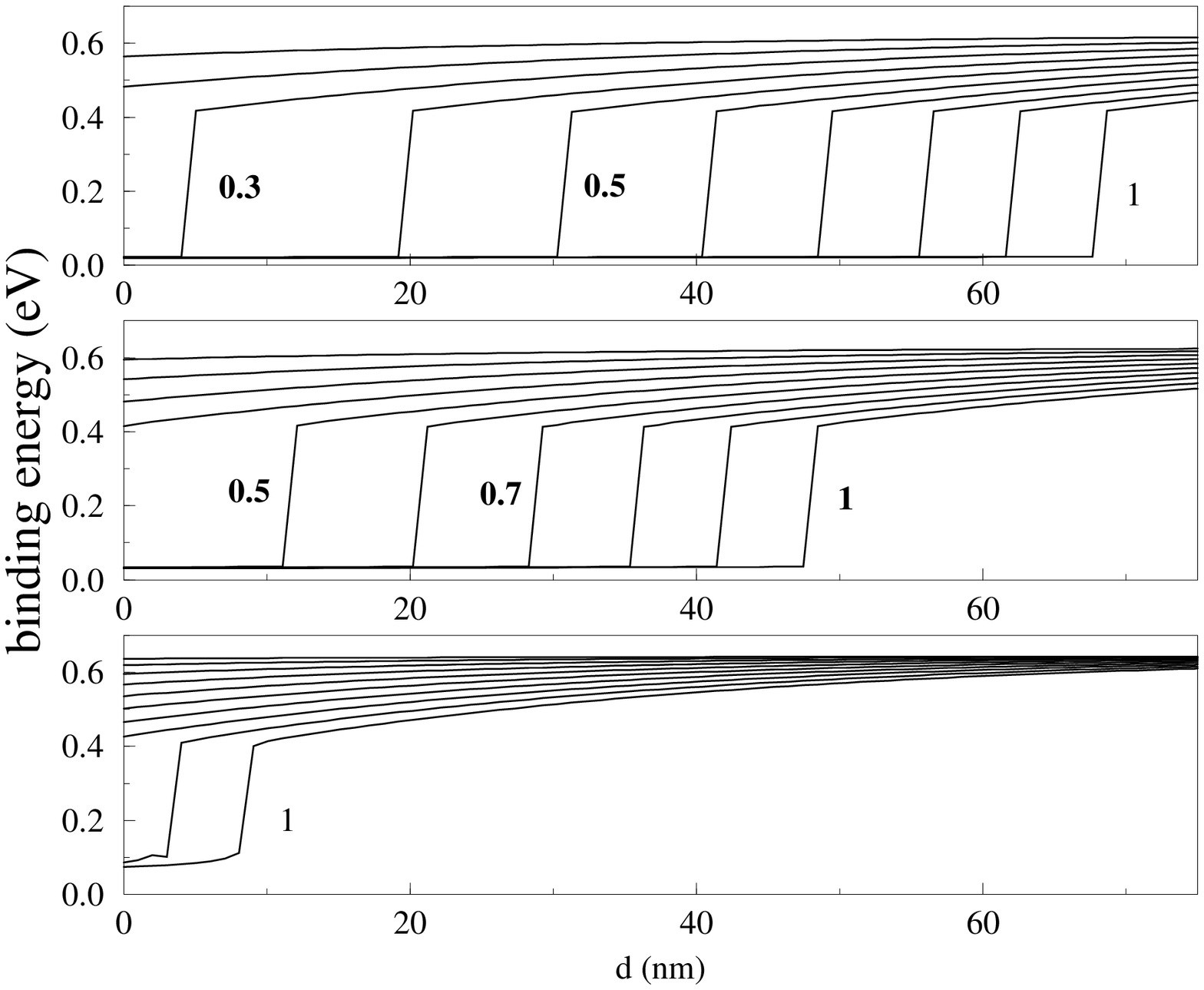}
\caption{V. Narayan PRL
\label{fig1}}  
\end{center}
\end{figure}

Fig~\ref{fig5} shows the instantaneous acceleration $a$  as a function
$d$ and $t$ for nanotubes with lengths $L=20-50$~nm. Comparing with
Fig~\ref{fig1} $a$ increases by several orders of magnitude  when the
exciton disassociates and the electron and hole reside at opposite
ends of the nanotube. The motion of a nanotube will be hindered by
collisions with the molecules of the liquid and will be strongly
dependent on the nanotube length.  The dependence of the nanotube
motion with length, can be the basis of its seperation according to
length by a thermal ratchet mechanism\cite{oud}.  The result indicate
that exciton disassociation can be achieved using modest electric
field gradients.  The field may have to further increased however to
overcome the random Brownian motion of the nanotube.

We have suggested a method to sort semiconducting nanotubes using
dielectrophoresis\cite{Krupke} and spectroscopy
techniques\cite{ostojic}. Since both of these methods have been
separately applied to nanotubes, it should be relatively straight
forward to combine them within a single experiment.  The experiments can
be accurately modeled by including the  exciton creation and
annihilation rates\cite{bunning} and the interaction with the liquid
by molecular dynamics simulations\cite{English}.  Recently, solar
cells\cite{sargent} have been made from polymers-quantum dots
composites. The quantum dots formed  in solution can be separated
according to their band-gap using light induced dielectrophoresis to
optimize the solar cell. We anticipate that future devices can be self
assembled by carefully fabricating electrodes and to
define electric field lines\cite{hunt} that will guide nano-ojects to
the required place. We hope this paper will stimulate future
experimental and theoretical work.

\bibliography{tube.bib}

\clearpage
\section*{Figure Captions}
\subsection*{Figure 1}
The binding energy as a function of $d$ and $t/T$ for
$L=50$~nm (top), $L=35$~nm (middle) and $L=20$~nm (bottom).
The lines are shown for $t/T=0.1-1.0$ in steps
of $0.1$, and some lines are labelled.  

\subsection*{Figure 2}
The electron (top) and hole (bottom) charge densities for a nanotube
with length $L=30$~nm, for $d=39.15$~nm (full lines), $d=39.05$~nm
(dotted lines), and $d=37.65$~nm (dashed lines).  The lines are shown
for $t/T=1$.
\subsection*{Figure 3}
The binding energies as a function of $t$ for the ground state (dotted
line) and first excited  state (full lines). Here $L=50$~nm,
$d=25$~nm,  $m_{e}=m_{h}=0.08 m_{0}$ for the ground state, and
$m_{e}=0.05 m_{0}$ and $m_{h}=0.08 m_{0}$ for the excited state.
\subsection*{Figure 4}
The dipole moment of a nanotube with length $L=50$~nm as a function of
an applied linear electric field.  The polarizability is the gradient.
\subsection*{Figure 5}
The log of the instantaneous acceleration as a function of $d$ and
$t/T$ for $L=50$~nm (top), $L=35$~nm (middle) and $L=20$~nm (bottom).
The lines are shown for $t/T=0.1=1.0$ in steps of $0.1$, and
some lines are labelled.

\clearpage

\clearpage
\begin{figure}[t!]
\begin{center}
\includegraphics[width=8cm,angle=0]{fig2.eps}
\caption{V. Narayan PRL
\label{fig2}}  
\end{center}
\end{figure}

\clearpage
\begin{figure}[t!]
\begin{center}
\includegraphics[width=8cm,angle=0]{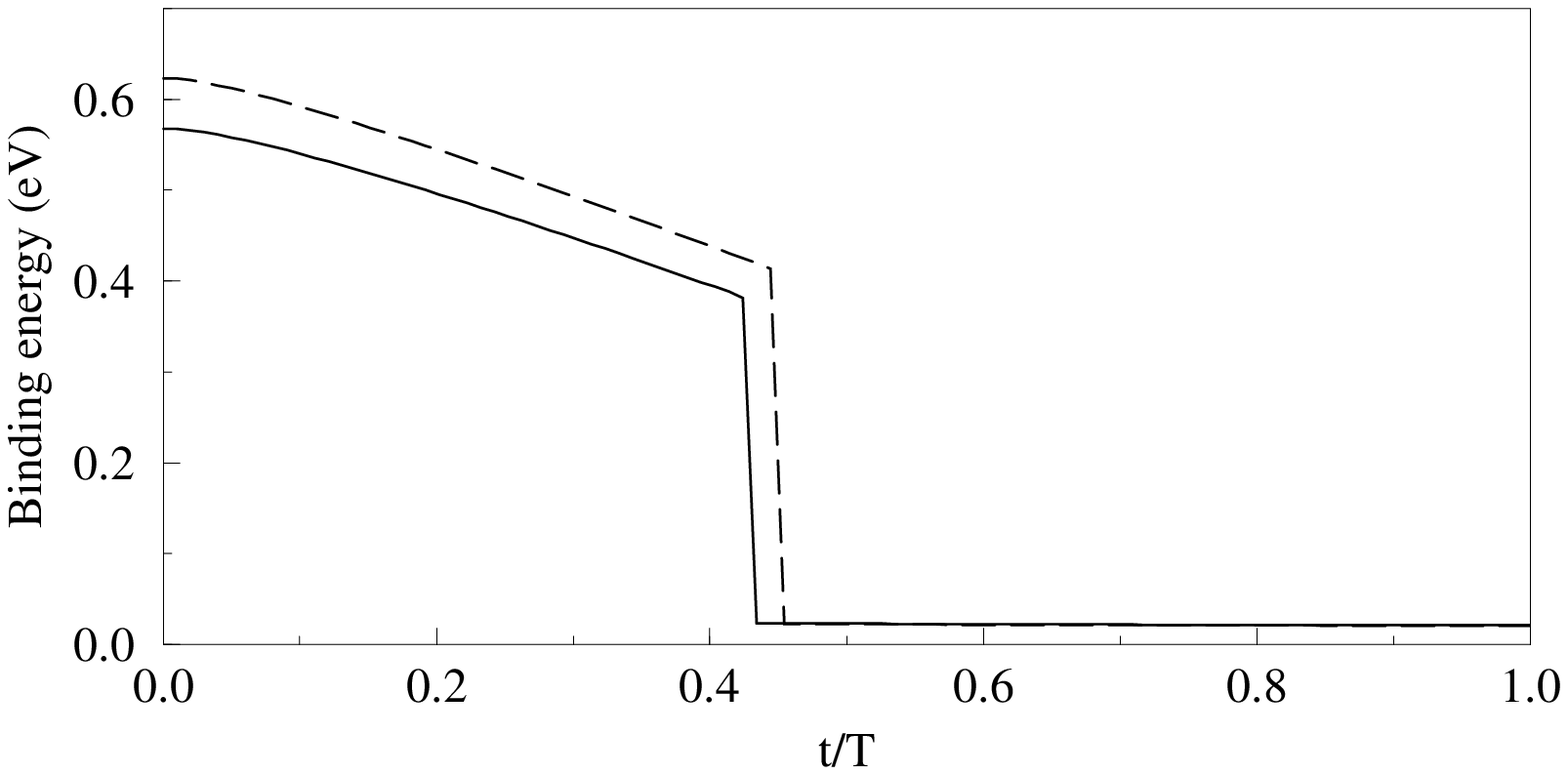}
\caption{V. Narayan PRL
\label{fig3}}  
\end{center}
\end{figure}

\clearpage
\begin{figure}[t!]
\begin{center}
\includegraphics[width=8cm,angle=0]{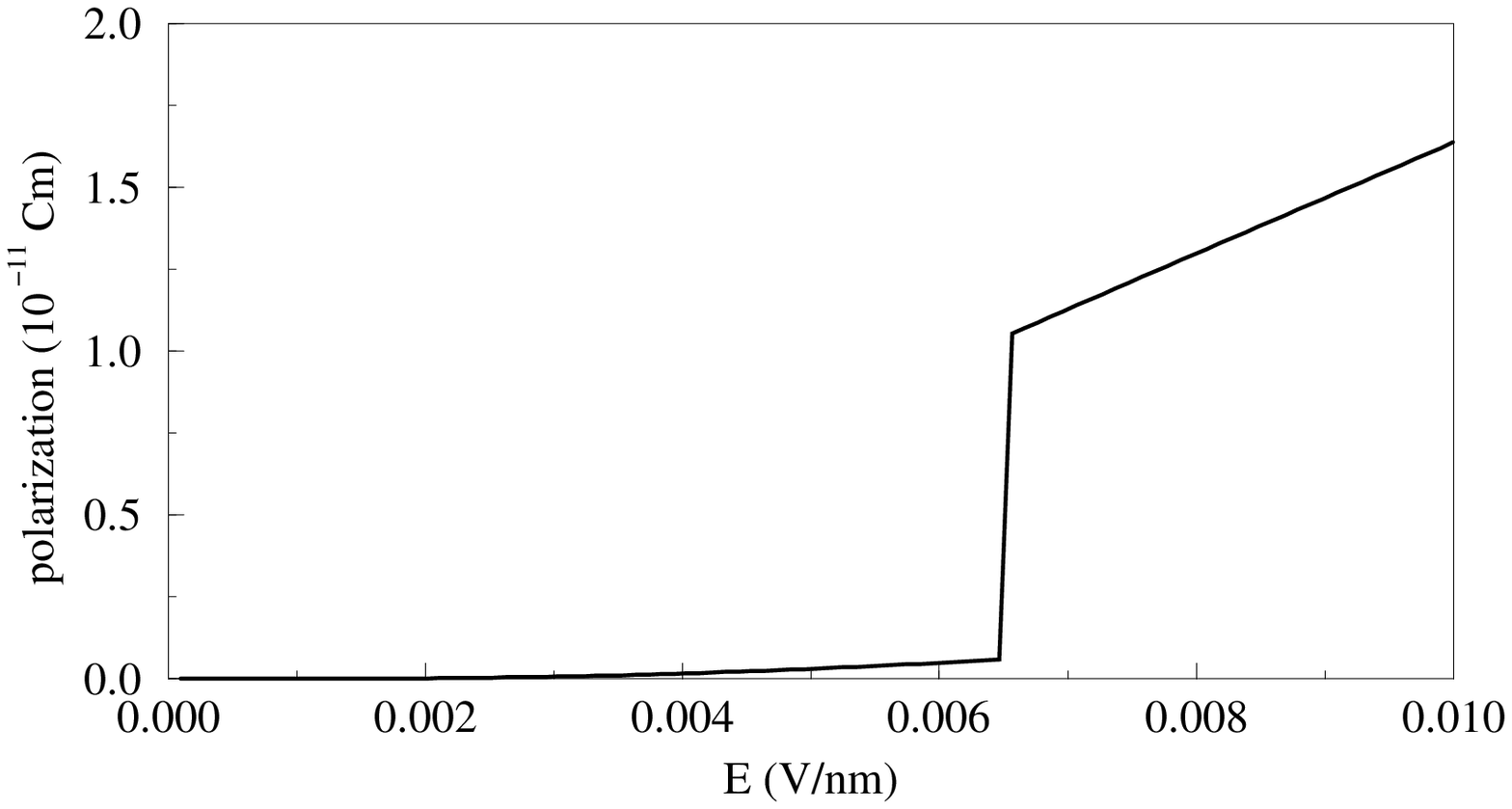}
\caption{V. Narayan PRL
\label{fig4}}  
\end{center}
\end{figure}

\clearpage
\begin{figure}[t!]
\begin{center}
\includegraphics[width=8cm,angle=0]{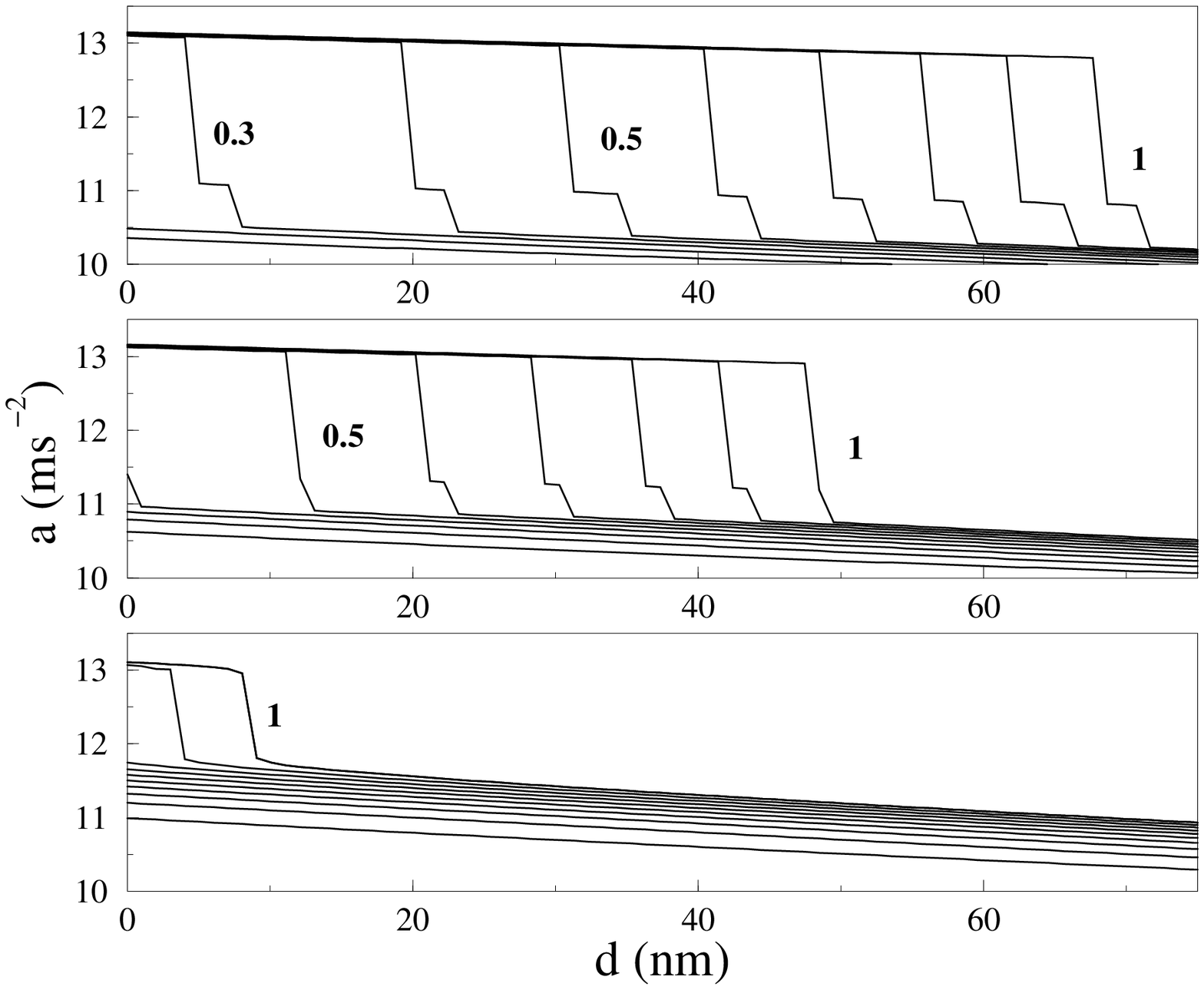}
\caption{V. Narayan PRL
\label{fig5}}  
\end{center}
\end{figure}

\end{document}